\documentclass[13pt,a4paper,english]{article} 
\usepackage[top=1in, bottom=1.25in, left=1in, right=1in]{geometry}
\usepackage{babel}   
\usepackage{url}     
\usepackage{hyperref}
\hypersetup{
    colorlinks=true,
    linkcolor=blue,
    filecolor=magenta,      
    urlcolor=blue,
    pdftitle={Overleaf Example},
    pdfpagemode=FullScreen,
    }
    \usepackage[T1]{fontenc}
    
    \usepackage{setspace}
\usepackage{indentfirst}
\usepackage{amsmath}                                    
\usepackage{amssymb}                                    

\usepackage{graphicx}
\usepackage{caption}
\usepackage[labelformat=empty]{caption}

\graphicspath{ {./images/} }
\title{%
  \bf BRUNO TOUSCHEK REMEMBERED\footnote{On December 2-3-4, 2021, a Memorial Symposium  to celebrate the 100th anniversary of the birth of Bruno Touschek,  was held in three different locations, corresponding to the institutions where Touschek gave major scientific contributions (links to the conference are provided here for each location): the \href{https://youtu.be/GwLpnoYRJKg}{Physics Department of Sapienza University of Rome}, the \href{https://www.youtube.com/playlist?list=PLRuUrPCVPFIquevS6-43fI1wwFVsRuC9_}{INFN Frascati National Laboratories} and the \href{https://www.lincei.it/it/manifestazioni/bruno-touschek-memorial-symposium}{Accademia Nazionale dei Lincei}.}
  \\  \textcolor{red}{1921 -- 2021} \\ 
 Bibliography and Sources}

\author{Luisa Bonolis\\Max Planck Institute for the History of Science, Berlin, Germany\\lbonolis@mpiwg-berlin.mpg.de}

\begin{document}
\maketitle


\abstract{The 100th anniversary of Bruno Touschek’s birth also marks 60 years since the first beams of electrons and positrons circulated in AdA, the first ever matter-antimatter collider built in Frascati National Laboratories following Touschek’s visionary proposal of February 1960. A brief biography, an extensive bibliography, and a description of archives containing documents related to the life and science of the father of electron-positron physics are presented.}

\vskip 0.7 cm

\begin{figure}[h]
\includegraphics[width=7.9cm]{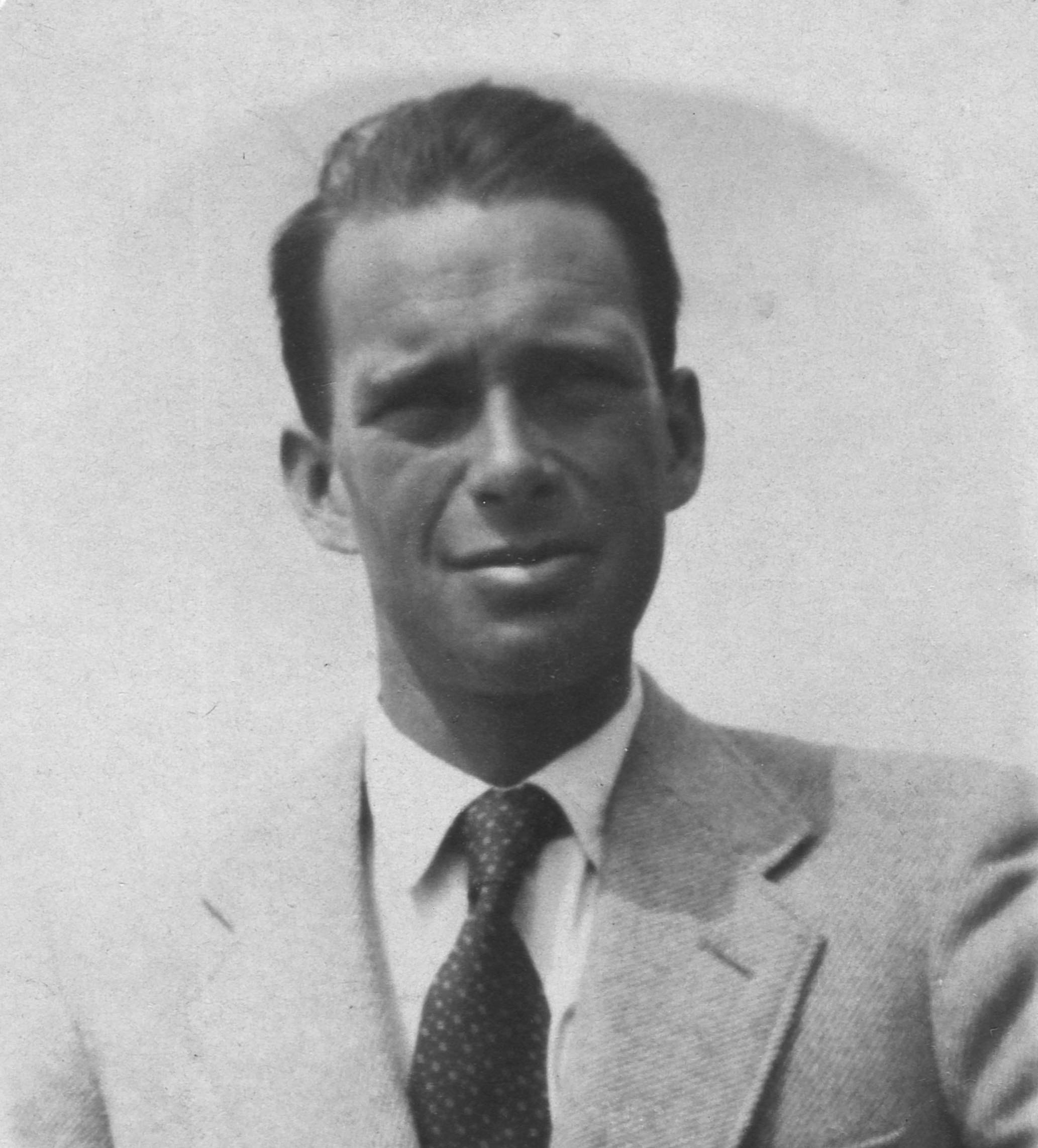}
\centering
\caption{Bruno Touschek in the 1950s. Courtesy of Francis Touschek.}
\end{figure}


\newpage

\section*{A brief biographical sketch}

Like many of his generation, Bruno Touschek went through a dramatic period of the last century, but he also experienced the enthusiasm and excitement of fighting for the reconstruction and revival of European physics after the tragedy of World War II. And indeed, his life both as a scientist and an intellectual, unfolded across Europe in space and time in different phases, moving either out of necessity, or driven by the desire for new challenges.

Touschek was born to Camilla Weltmann and Franz Xaver Touschek on February 3rd, 1921, in Vienna, where he spent his childhood and early youth. Between the end of the 19th century and the beginning of the First World War, the Austrian capital was a highly cosmopolitan city and one of the most important centers of scientific advancement. But it was equally a  center of creation for modernity, the cradle for a number of ideas which shaped the whole 20th century and flourished in art, architecture, design, literature, science, philosophy and music. At the epicenter of this multifaceted world was the writer and essayist Karl Kraus. His caustic satirical spirit and his cultural engagement in the fundamental ideological issues of his time, had a profound influence on Touschek’s intellectual formation.
The Vienna Circle, a discussion group of brilliant scientists, philosophers and mathematicians whose intellectual endeavor aimed to redraw the scientific conception of the world, was a further typical expression of the rich and radical intellectual culture of inter-war Vienna.

Touschek’s own family was actively involved in this scenario. His mother Camilla Weltmann and his aunt Ella -- and Ella’s own husband, the architect Josef Margold -- were active in the circle of the Wiener Werkst\"atte, the association evolving from the Vienna Secession founded by Josef Hoffmann and Koloman Moser as an alliance of artists, architects, designers and artisans that pioneered modern design and eventually influenced the Bauhaus movement and the Art Deco style. Touschek grew up in such great avant-garde cultural movements. 
In particular, his precocious talent for drawing, was influenced by the innovative expressionism of Egon Schiele and the famous psychological portraits of Oskar Kokoschka.
The style of his own drawings – well-known among friends and colleagues – testifies the persistence during his whole life of such strong and lively bonds with the rich cultural and intellectual world of his home town, that subtly blended in his personal and very original style.

However, due to his Jewish origin on the maternal side, the annexation of Austria into Nazi Germany in 1938 completely turned his life upside down and dramatically affected his future forever. He was no longer able to attend classes as a regular student at school and in spring 1939 he was forced to take his Matura, his final examination, privately. 
He was grown up enough to be deeply aware of and suffer from the dramatic events that were happening around him.

Touschek was able to enroll in Physics at the University of Vienna, but again, in June 1940, he was expelled from the University for racial and political reasons and could only continue his studies privately, helped by Paul Urban, who had received his PhD in theoretical physics under the supervision of Hans Thirring. Urban put him in contact with Arnold Sommerfeld, one of the oldest and best-loved representatives of German physics, who 
had educated and mentored a whole generation of young physicists and students (notably Werner Heisenberg and Wolfgang Pauli), who had in turn a key role in the new era of theoretical physics. 
Helped by Sommerfeld, in early 1942 Touschek abandoned Vienna and with great courage continued to pursue his passion for physics during the war years. Despite still being unable to attend classes as a regular student, he continued his studies in Germany protected by Sommerfeld’s colleagues and friends in Hamburg and later in Berlin, and at the same time worked to support himself. 

  During those difficult times, he had the chance to come in contact with some of the most influential European physicists, and found himself involved in the construction of a 15-MeV betatron, a pioneering accelerator designed by the Norwegian scientist Rolf Widerøe. He also went through extremely distressing experiences, such as being imprisoned by the Gestapo and being shot during a march towards the Kiel concentration camp. 
  
  He miraculously survived such dreadful events and after the war, finally obtained his Diploma in Physics at the University of G\"ottingen with a dissertation on the theory of the betatron and for some time was Werner Heisenberg’s assistant at the Max Planck Institute for Physics, continuing his formation under the influence of the great German theoretical school. 
  
  Despite having lost several years of his youth, he re-emerged from the war and early-post war years as one of the first physicists in Europe endowed with a unique expertise about the theory and functioning of accelerators. Such difficult period in his life proved to be a major step along his path to the conception of the first ever matter-antimatter collider which he eventually proposed and built in Frascati National Laboratories in 1960.

In early 1947 Touschek moved to Glasgow University and in 1949 was awarded a PhD in Physics with a dissertation on “Collisions between electrons and nuclei” (with Rudolf Peierls as doctoral supervisor), while being involved in theoretical studies and in the building of a 300-MeV electron synchrotron, also consulted as a betatron expert by other research centers in UK. In Glasgow, he became a full-fledged theoretical physicist. At that time, he came in contact with Max Born, one of the fathers of the new quantum mechanics, who had emigrated to UK after having been suspended from his position at the University of G\"ottingen being of Jewish origin. Touschek collaborated with him writing an appendix for a new edition of Born’s book {\it Atomic Physics}. 

 \begin{figure}[h]
\includegraphics[width=6.5cm]{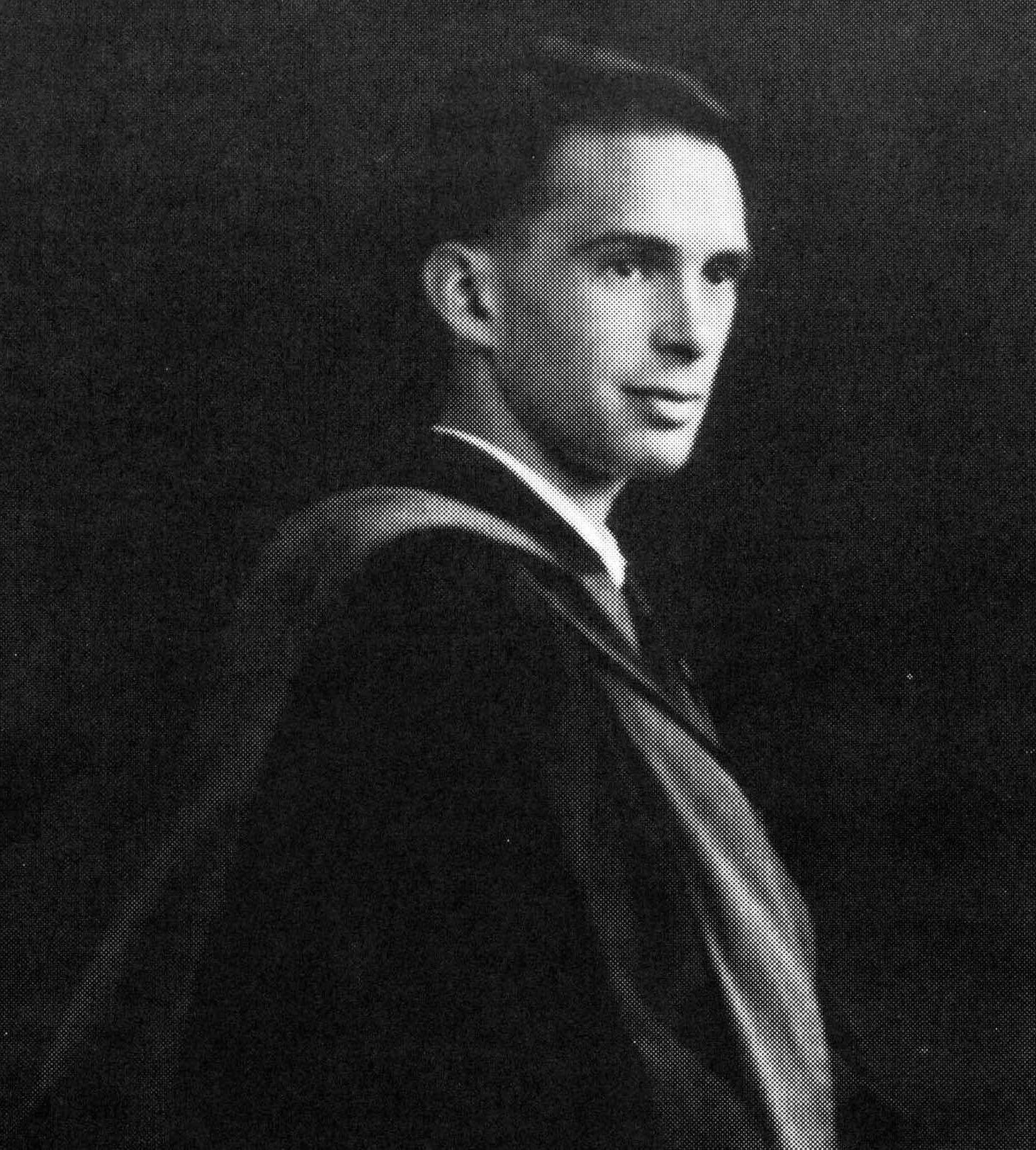}
\centering
\caption{Bruno Touschek in 1949. Courtesy of Francis Touschek.}
\end{figure}

In the very early 1950s, Touschek met the well-known Italian theoretical physicist Bruno Ferretti, with whom he shared lively discussions on quantum field theory and from whom he also heard about the great plans and expectations for the reconstruction and revival of Italian physics, after the disaster of World War II. Edoardo Amaldi and Gilberto Bernardini, both heirs of the scientific tradition established by Enrico Fermi and Bruno Rossi in the 1930s, were in complete harmony in their efforts with the scientific community to restore the pre-war excellence of Italian and European physics. Amaldi, in particular, was one of the promoters and founding fathers of CERN, where Bernardini soon became Director of Research. On the other hand, as we know from letters to his father and especially to Arnold Sommerfeld, since some time Touschek felt that in Glasgow he was rather far from the mainstream of theoretical physics and was considering other possible positions in Europe. Full of enthusiasm about the idea of Italy, a country he knew since his childhood, when he visited his maternal aunt Ada married with an Italian, he decided to obtain a leave of absence and move to Rome. Amaldi, deeply aware of Touschek’s potential, invited him officially with an INFN contract.

 \begin{figure}[h]
\includegraphics[width=7.5cm]{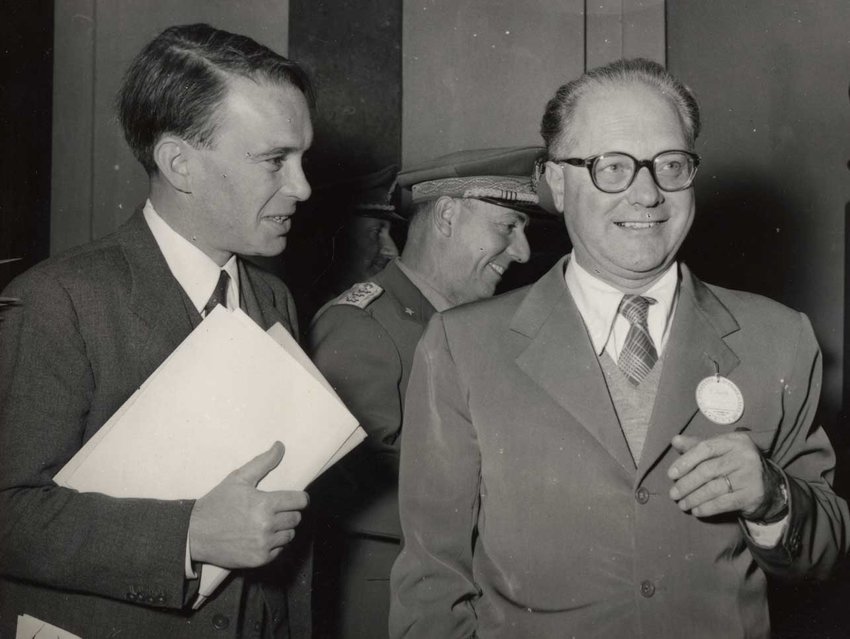}
\centering
\caption{Touschek and Edoardo Amaldi in the 1960s. Courtesy of Francis Touschek.}
\end{figure}

When Touschek arrived in Italy at the end of 1952, the Istituto Nazionale di Fisica Nucleare had just been founded and Italian physicists were deciding to build an electron synchrotron, a new-generation machine which would become a powerful tool for high-energy physics. After having been for many years at the frontier of nuclear and cosmic-ray research, Italy hoped to regain a prominent position in the sub-nuclear realm and indeed Touschek’s unique expertise was destined to have a profound influence on the future of this field in Italy, both theoretically and experimentally. In the following years, he continued to play a role in the process of reconstruction and revival of physical sciences in Europe, and further evolved as a theoretician, giving relevant contributions to the study of discrete symmetries in particle physics and in neutrino physics. 
During the 1950s, while being actively involved in the life of the Italian scientific community, fitting brilliantly into its rich and dynamical academic and scientific environment, Touschek closely experienced the birth and development of Frascati National Laboratories that had been established to host the brand new accelerator. But between the end of 1959-early 1960, when the 1100 MeV electron synchrotron had just gone on line, Touschek surprised everybody proposing to go far beyond experiments with gamma beams obtained by hitting electrons against a fixed target inside the synchrotron, or even experiments such as those US physicists were scheduling at Stanford with two colliding beams of electrons stored in two tangent rings.\footnote{D. W. Kerst et al. 1956. \href{https://doi.org/10.1103/PhysRev.102.590}{Attainment of Very High Energy by Means of Intersecting Beams of Particles}; G. K. O'Neill. 1956. \href{https://doi.org/10.1103/PhysRev.102.1418}{Storage-Ring Synchrotron: Device for High-Energy Physics Research}. G. K. O' Neill and J. A. Ball, U. S. 1957. Atomic Energy Commission Report No. US AEC NYO-8015; G. K. O Neill. 1958. Bull. Am. Phys. Soc. 3, 170; W. C. Barber, B. Gittelman, G. K. O' Neill, W. K. H. Panofsky, and B. Richter. 1959. Stanford University High Energy Laboratory Report No. HEPL 170, 46; G. O’Neill. 1959. \href{https://inspirehep.net/literature/919845}{Storage Rings For Electrons And Protons}. In L. Kowarski (ed) Proceedings, 2nd International Conference on High-Energy Accelerators and Instrumentation, HEACC 1959: CERN, Geneva, Switzerland, September 14-19, 1959, 125-133; G. O’Neill. 1959. \href{http://inspirehep.net/record/919833}{Experimental methods for colliding beams}. In L. Kowarski (ed.) Proceedings,2nd International Conference on High-Energy Accelerators and Instrumentation, HEACC 1959: CERN, Geneva, Switzerland, September 14-19, 1959, 23-25. Their final results were published in 1966: W. C. Barber et al. 1966. \href{https://doi.org/10.1103/PhysRevLett.16.1127}{Test of Quantum Electrodynamics by Electron-Electron Scattering}.}  
According to Touschek, what would really be worth exploring, instead, was the physics of electron-positron annihilations, which would allow to open a channel into the hadronic world through the quantum numbers of $e^+e^-$. The colliding beam technique, which other physicists were planning to exploit both in USA and USSR -- basically to obtain a larger center-of-mass energy or to perform high-precision experiments to test the predictions of QED -- was definitely moving towards a conceptually novel stage. Moreover, because of the CPT symmetry, an $e^+e^-$ machine could be realized with a single magnetic ring, ensuring that electrons and positron could circulate in the same orbit, in opposite directions,  and eventually collide. 

 \begin{figure}[h]
\includegraphics[width=9.5cm]{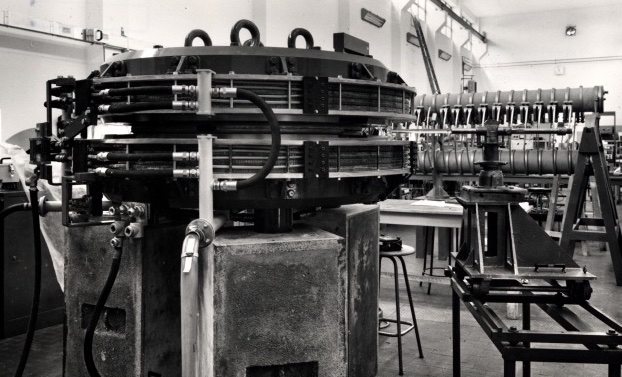}
\centering
\caption{In about one year AdA, the first $e^+e^-$ storage ring, was ready for beam tests on a gamma beam line of the synchrotron. Electrons and positrons were injected in the ring by conversion on an internal target. Courtesy of INFN/LNF.}
\end{figure}


Following his challenging ideas – based on his firm belief in CPT and QED – the first matter antimatter collider AdA (for ‘Anello di Accumulazione’, Storage Ring) was built, inaugurating a brand-new research line at Frascati National Laboratories and heralding a new era in high-energy physics. The team led by Touschek, including Carlo Bernardini, Giorgio Ghigo, Gianfranco Corazza, Ruggero Querzoli and Giuseppe Di Giugno, was able already on February 1961 to observe the light signal from a single circulating electron after its capture in the ring as a pulse in the phototube output, or even as a white-bluish spot that could be seen with the naked eye through a small porthole. 

 In Orsay, at the Laboratoire de l’Acc\'el\'erateur Lin\'eaire, where AdA was moved in 1962 to benefit from the Linac high particle injection rates,  Touschek and his team, now including Pierre Marin and Jacques Ha\"issinski, were able to demonstrate that the two beams had actually collided, thus proving the feasibility of such machines.\footnote{C. Bernardini, G. Corazza, G. Giugno, J. Ha\"issinski, and P. Marin. \href{https://doi.org/10.1007/BF02750550}{Measurements of the rate of interaction between stored electrons and positrons. Il Nuovo Cimento, 34(6): 1473-1493, December 1964.}} They also discovered an unexpected effect, a loss of particles from the stored beams reducing their lifetime, whose origin was immediately explained by Touschek as an intra-bunch scattering. The so-called {\it Touschek effect} luckily scales sharply with energy and does not seriously affect more powerful colliders.\footnote{C. Bernardini, G. F. Corazza, G. Di Giugno, G. Ghigo, R. Querzoli, J. Ha\"issinski, P. Marin, and B. Touschek. \href{https://doi.org/10.1103/PhysRevLett.10.407}{Lifetime and beam size in a storage ring. Physical Review Letters, 10(9): 407-409, May 1963.}} 
 
  \begin{figure}[h]
\includegraphics[width=6.5cm]{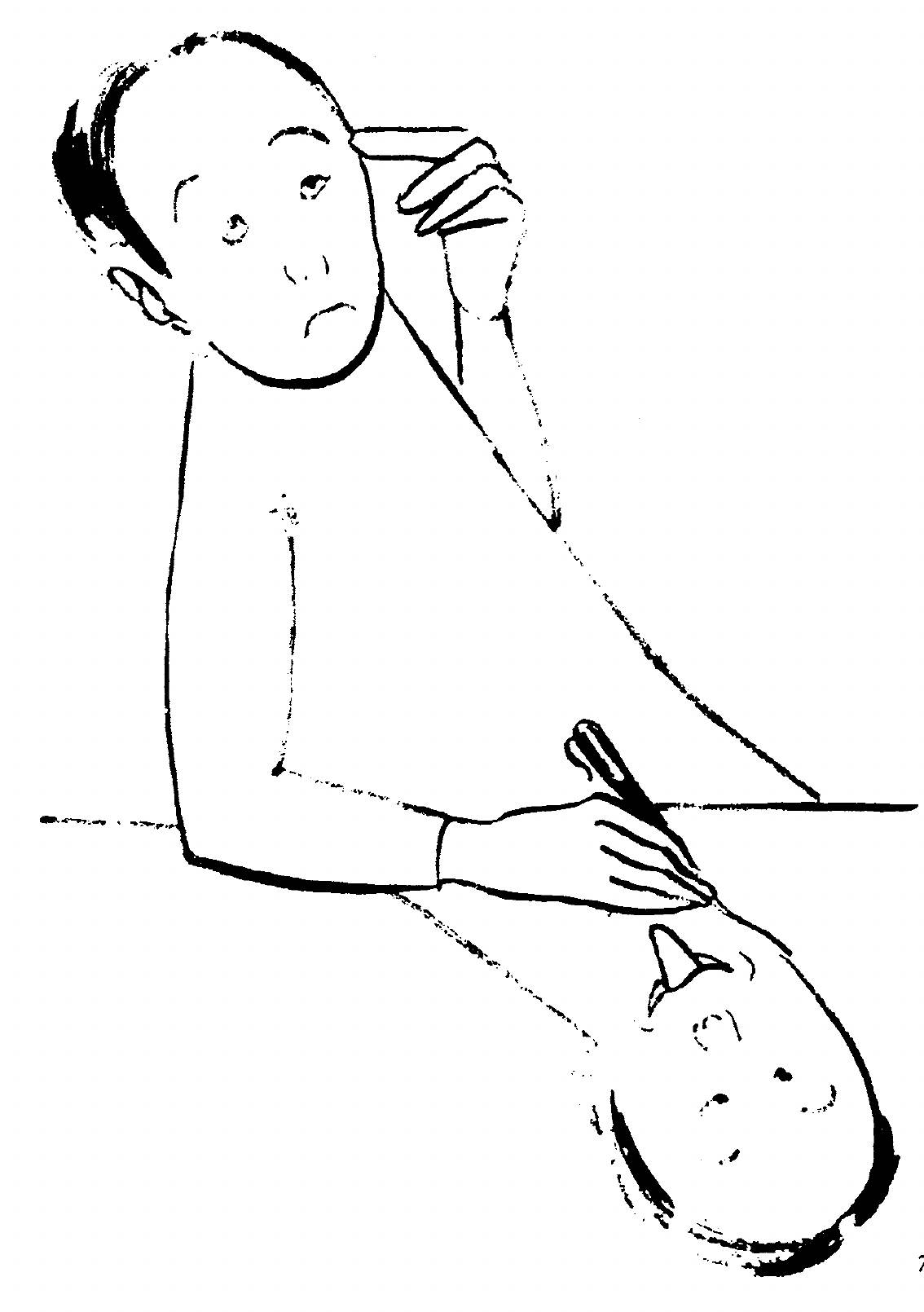}
 \centering
\caption{Bruno Touschek, ironic portrayal of T. D. Lee and parity violation. Courtesy of Francis Touschek.}

\end{figure}

In November 1960, even before AdA had showed the feasibility of electron-positron collisions, opening the way to higher energy and luminosity, Touschek had prepared a draft plan for ADONE, a larger and more powerful electron-positron collider. 

 \begin{figure}[h]
\includegraphics[width=7.7cm]{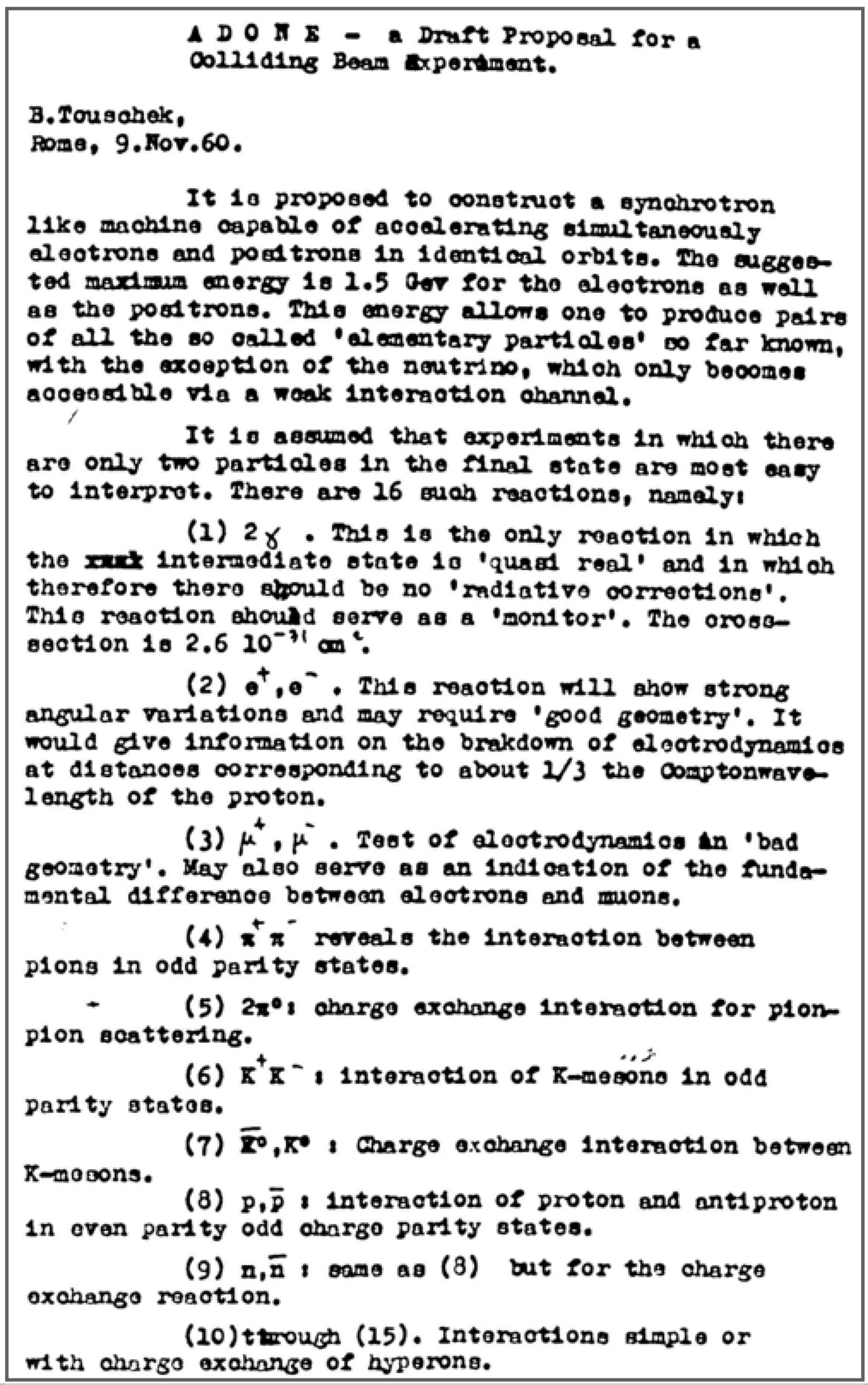}
 \centering
\caption{Bruno Touschek: first draft proposal for ADONE, November 9, 1960. Courtesy of `Edoardo Amaldi' Archives, Physics Department, Sapienza University of Rome.}

\end{figure}

ADONE (1.5 GeV per beam, 105 m in circumference), which became operational in 1969, eventually discovered the multi hadron production making electron-positron physics a field of major interest, further encouraging the construction of a large family of high-energy colliders all over the world where new types of elementary constituents of matter were detected.
In 1974, ADONE confirmed the existence of the $J/\psi$, a bound state of a charm quark and a charm anti-quark, discovered at the Brookhaven National Laboratory and at the Stanford Linear Accelerator Center with the $e^+e^-$ collider SPEAR. It was the first breakthrough discovery of the colliding beam technique and the first firm experimental evidence for the charm quark -- establishing the quark model as a credible description of nature -- for which Burton Richter and Samuel Ting were awarded the 1976 Nobel Prize in Physics.

With AdA and ADONE Touschek created a brand new, major research line at Frascati Laboratories and made a further fundamental contribution with the formation of a theoretical school in Rome and Frascati. In particular, he had been Nicola Cabibbo’s thesis advisor in 1958. While AdA was being built, Cabibbo and Raoul Gatto’s thorough theoretical analysis of the annihilation processes of interest  resulted in a seminal paper universally known as the ‘Bible’. Touschek, who was particularly valued also as an extremely brilliant and fascinating teacher, in 1972 was elected as Foreign Member of the Accademia Nazionale dei Lincei, while in 1975 was awarded the prestigious Matteucci Medal by the Italian National Academy of Sciences.

Between 1977-1978, Touschek spent the last months of his life as visiting scientist at CERN, at a time when early plans for a giant electron-positron collider were being discussed. However, when LEP, with an energy of 50 GeV per beam and a circumference of about 27 km, eventually came into operation in 1989, Touschek was no more there. He had prematurely passed away, on the 25th of May 1978, while he was participating in the planning of the proton-antiproton collider ($Sp\bar p S$) proposed by Carlo Rubbia. Unfortunately, he did not live enough to witness the discovery of the $W^{\pm}$ and $Z$ bosons and the related 1984 Nobel Prize in Physics to Rubbia and Simon van der Meer. 

 \begin{figure}[h]
\includegraphics[width=7.5cm]{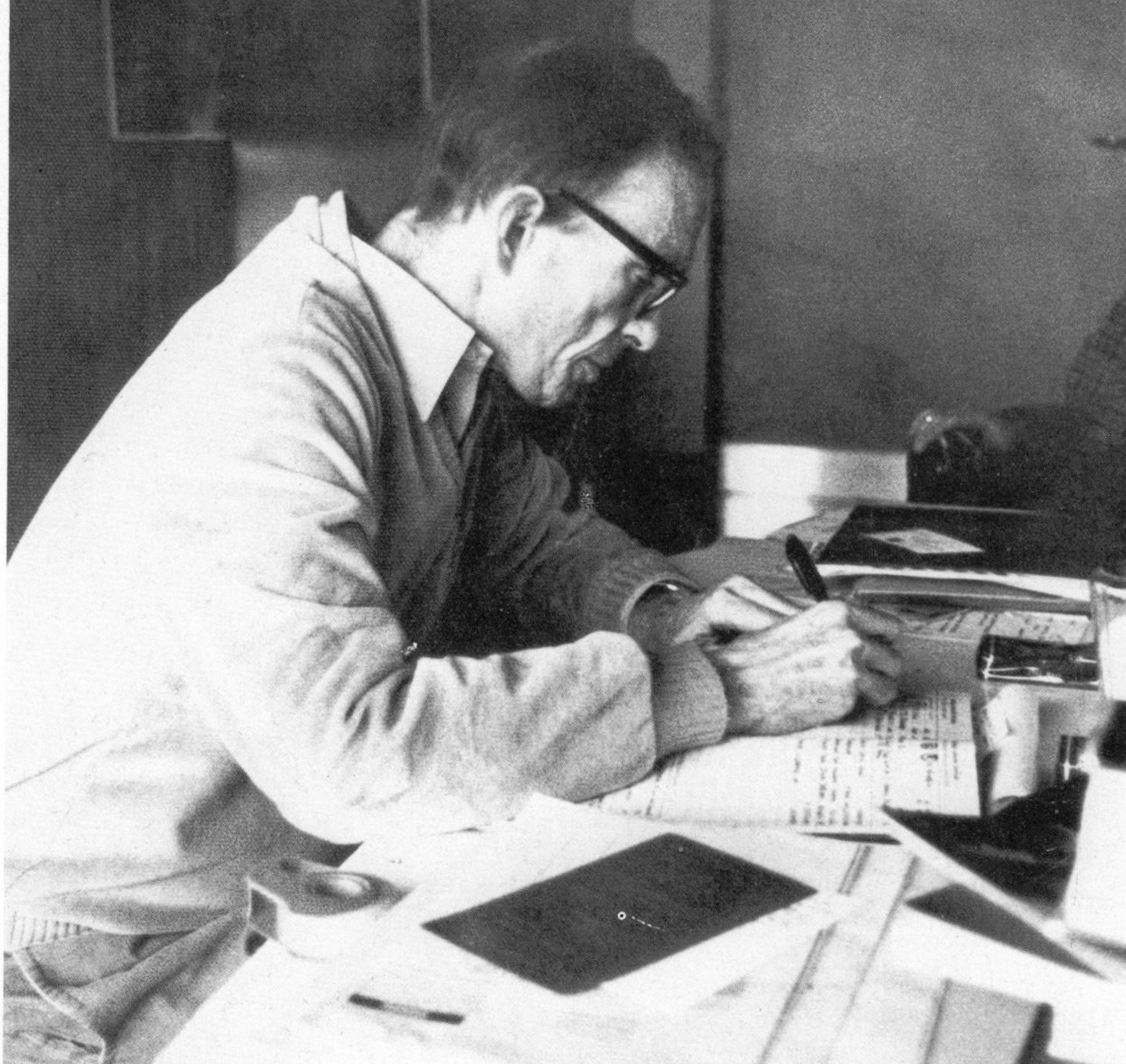}
\centering
\caption{Bruno Touschek during the last months of his life. Courtesy of Francis Touschek.}
\end{figure}

Bruno Touschek’s small AdA (about 1.3 meters in diameter, storing beams of 250 MeV) has opened the way to new bigger matter-antimatter colliders and precision measurements which have been instrumental in confirming our understanding of the basic building blocks of matter in the Universe and the fundamental forces that operate between them. The detection of the long-sought Higgs boson at the Large Hadron Collider at CERN in 2012, has eventually completed the Standard Theory of particle physics. The fundamental contribution of AdA as a progenitor of entire generations of colliders was recognized on 5 December 2013, when the world’s first particle-antiparticle accelerator – still visible on the grounds of INFN Frascati National Laboratories – was declared an Historic Site by the European Physical Society. This important recognition has definitely marked AdA’s role as a milestone in the Italian and European scientific heritage.
 
 \begin{figure}[h]
\includegraphics[width=6.1cm]{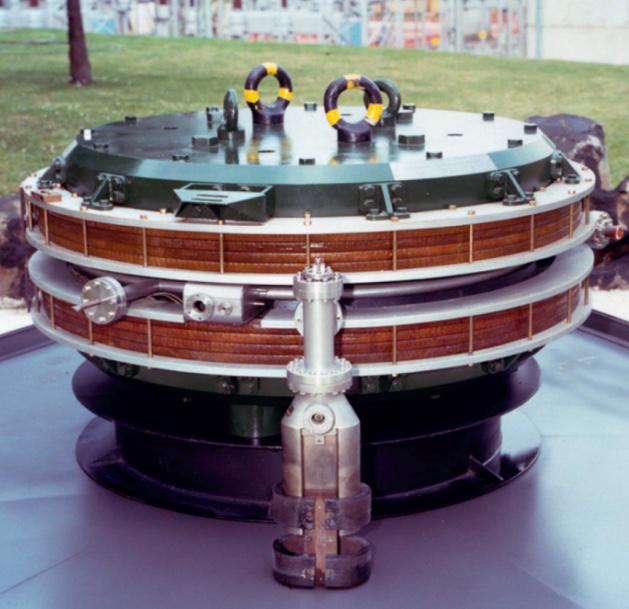}
\centering
\caption{AdA on display today at the EPS Historic Site on the grounds of Frascati National Laboratories. Courtesy of INFN/LNF.}
\end{figure}


\section*{Documents and Sources}

Immediately after his death, the first to write  on Bruno Touschek and AdA was his colleague and friend Carlo Bernardini [Bernardini 1978]. Soon after, Edoardo Amaldi  published  a long biography [Amaldi 1981], which includes a list of Touschek's papers as well as examples of his famous drawings. 

A fundamental source is the volume {\it Le carte di Bruno Touschek} describing in detail his personal papers preserved at the `Edoardo Amaldi Archives', Physics Department of Sapienza University in Rome [Battimelli, De Maria, Paoloni 1989]. Information on this collection, which can be consulted at the Library of the Physics Department, is also available online at \url{https://archivisapienzasmfn.archiui.com/oggetti/5-bruno-touschek}.

Other main archives containing documents related to Bruno Touschek are CERN Archives (correspondence with Wolfgang Pauli and L\'eon Van Hove), the Max Planck Society Archives in Berlin (correspondence with Werner Heisenberg), the Deutsches Museum Archives in Munich (correspondence with Arnold Sommerfeld and his son Ernst Sommerfeld), the University of Glasgow Archives \& Special Collections (documents related to his years in Glasgow), Churchill Archives Centre in Churchill College, Cambridge (correspondence with Max Born). Copies of reports on the theory of the betatron written by Touschek during and immediately after the war are preserved in Rolf Wider\o e's papers at the \href{https://www.research-collection.ethz.ch/bitstream/handle/20.500.11850/140811/eth-22301-01.pdf}{Eidgenossischen Technischen Hochschule} (ETH) in Zurich. The bulk of this material became part of his dissertation in G\"ottingen. Wider\o e's autobiography edited by Pedro Waloschek contains several pages mentioning Touschek and their relationship [Wider\o e 1994]. Waloschek's {\it Death-Rays as Life-Savers in the Third Reich} is also a valuable source on Bruno Touschek's life and the context he lived during the war [Waloschek 2012].

The establishment of the `Bruno Touschek Memorial Lectures' at INFN National Laboratories in Frascati -- promoted in 1987 by Giulia Pancheri, a former collaborator of Bruno Touschek --
 was marked by a conference whose proceedings include valuable contributions which can be consulted online [Greco, Pancheri 2004]. 
 
Several recollections by Touschek's colleagues are also contained in a volume edited by V. Valente ({\it Adone, a milestone on the particle way}, INFN, Frascati, 1997), but especially in a dedicated volume edited by G. Isidori ({\it Bruno Touschek and the birth of $e^+e^-$  physics}, Frascati Physics Series Vol. XIII, 1998).

\begin{figure}[h]
\includegraphics[width=9cm]{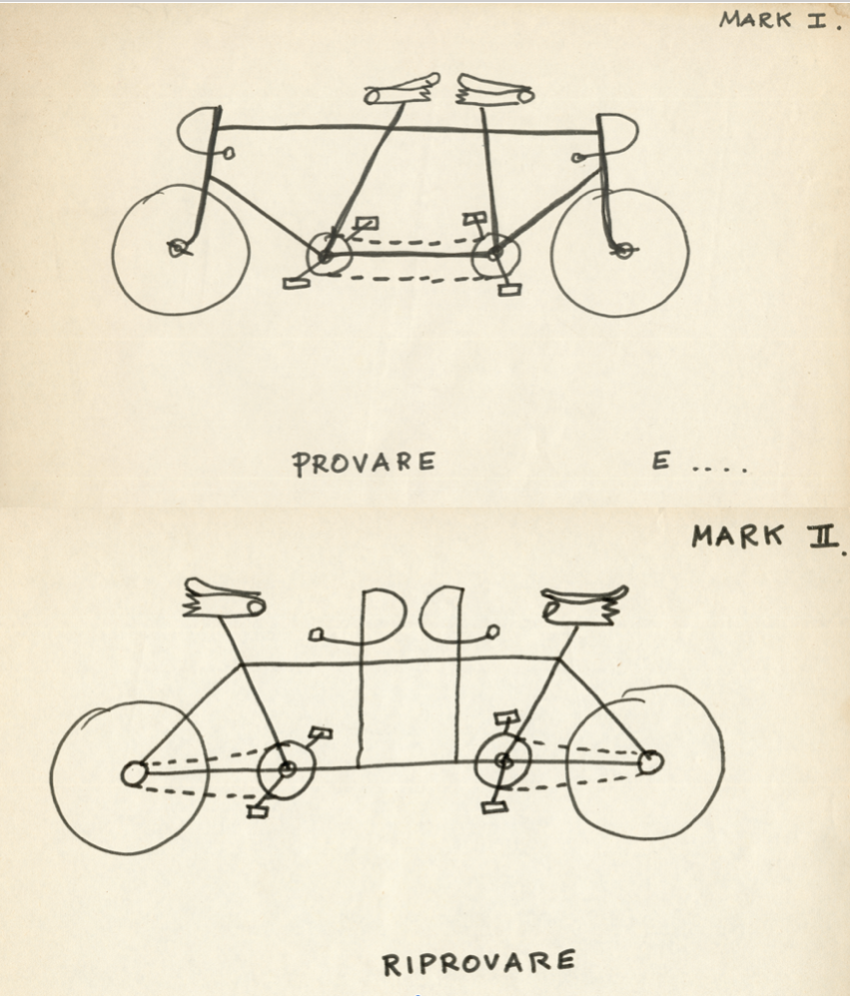}
\centering
\caption{One of Touschek's most famous drawings, showing his peculiar sense of humour, in which irony, satire, irreverent mockery, blended in a very subtle way. Courtesy of Francis Touschek.}
\end{figure}

More systematic historical studies began in the early 2000s, during the realization of the docu-film {\it Bruno Touschek and the Art of Physics} [Agapito, Bonolis 2004] also thanks to Touschek's widow, the late Elspeth Yonge, who kindly gave access to extremely relevant papers, correspondence and photos still preserved by the family. For early results of this research work see [Bonolis 2005a,b]. 

\begin{figure}[h]
\includegraphics[width=9cm]{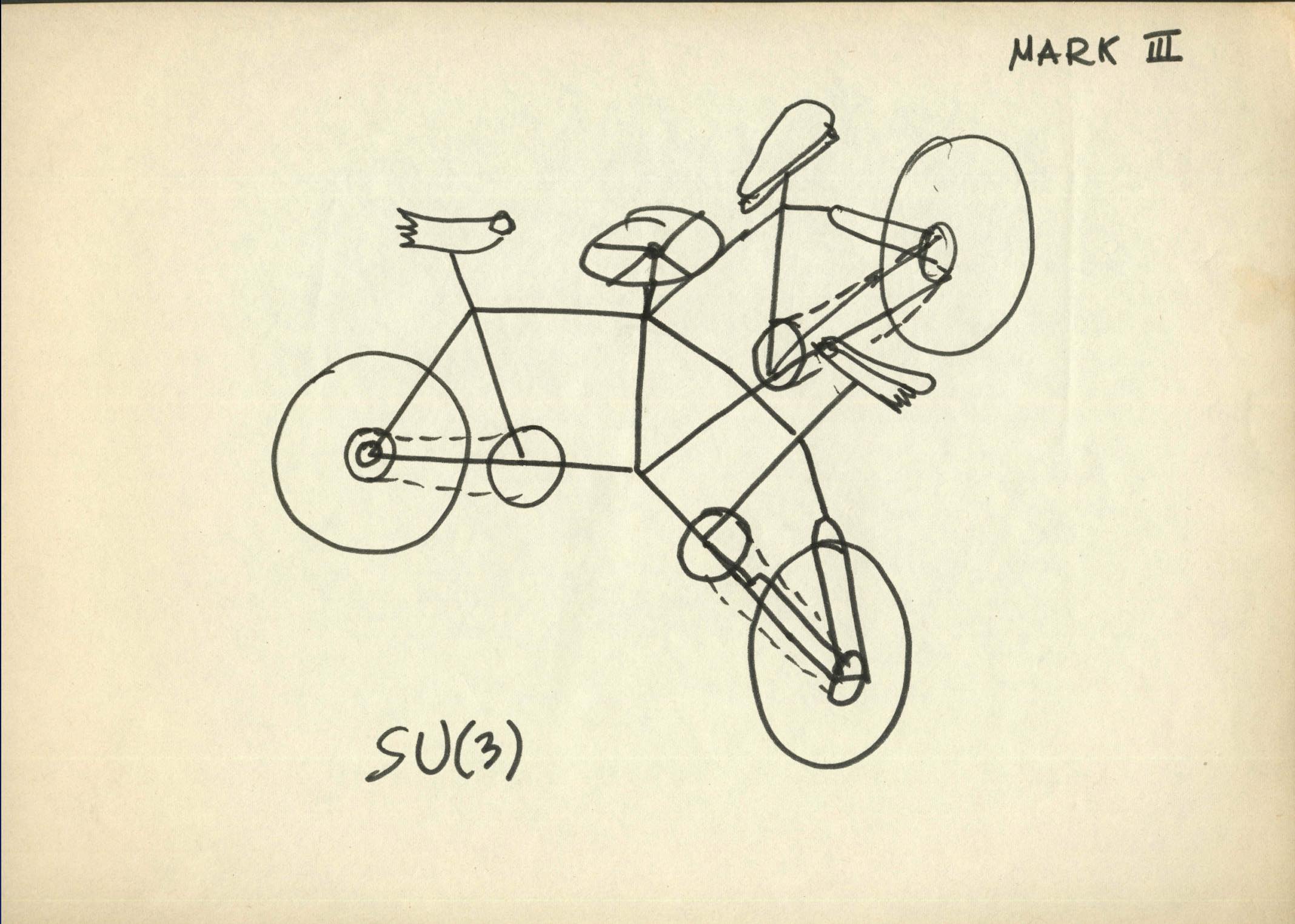}
\centering
\caption{The third drawing in the series of bicycles. Courtesy of Francis Touschek.}
\end{figure}

Since 2009, an in-depth historical analysis has been conducted by Bonolis and Pancheri resulting in several contributions reconstructing different phases of Bruno Touschek's life, especially emphasizing his formation both as a theoretical physicist and an expert in accelerator science and showing how the merging of such articulated competences became the origin of his visionary proposal to build the first electron-positron collider and investigate the matter-antimatter physics (see Bibliography). Further relevant documents, such as the collection of Touschek’s letters to his father put at disposal by Elspeth Touschek, have been instrumental for such research work, in particular because they have thrown light upon his crucial war and early post-war years, which had remained rather unknown apart from the very few episodes he had told Amaldi during the last days of his life. Part of the results of these studies were used  during the realization of a second docu-film: {\it Bruno Touschek with AdA in Orsay} [Agapito, Bonolis, Pancheri 2013].

A comprehensive biography of Bruno Touschek by Giulia Pancheri is under publication with the title {\it Bruno Touschek's Extraordinary Adventure -- from death-rays to 
anti-matter} (Springer).

\section*{Conclusion}

The centennial of Bruno Touschek's birth inaugurates a new phase in the historical studies on one of the most original figures of 20th century physics.
 As historians, we are still left with the task of in-depth investigations related to the evolution of  Touschek’s theoretical thought during the twenty years or so between the war years -- and his work on the betatron theory -- to the late 1950s, when such a long process finally materialized into his daring and drastic proposal, that he considered ``the future goal'' of Frascati Laboratories: transform the electron synchrotron, that had just begun to function, into an electron-positron collider and explore the physics of matter-antimatter annihilations. His bold idea was wisely and enthusiastically converted into the decision to build a dedicated small prototype, AdA, the first ever matter-antimatter machine, which in the early 1960s set the stage for a new era in particle physics. 

The period from Touschek’s arrival in Italy at the end of 1952,  to the end of the 1950s, during which he fully developed into a mature theoretical physicist dialoguing with prominent theoreticians of his time, has not yet been thoroughly studied. In particular his scientific production as well as his scientific correspondence with Heisenberg, which dates back to the early post-war period, and continued during the 1950s, has yet to be analyzed, as well as his exchange of letters with Wolfgang Pauli, himself born in Vienna from a prominent Jewish family, of whose work Touschek had always been an attentive follower since his early youth. They had an intense scientific correspondence during 1957-1958, at a time when much of Pauli’s work was still centered on quantum field theory -- and had already resulted in two fundamental pillars of the theory: the spin-statistics theorem and the CPT theorem -- but in particular when the shocking discovery of Parity violation in weak interactions was increasing interest in the discrete symmetry operations, the charge conjugation $C$ and time reversal $T$. Such dialogue with Pauli is also testified by their joint paper \href{https://doi.org/10.1007/BF02724849}{Report and comment on F. G\"ursey’s <<Group structure of elementary particles>>} published in 1959 as a contribution to the {\it International School of Physics ``Enrico Fermi''} (8th Course: ``Mathematical problems of the quantum theory of particles and fields'') when Pauli had already passed away. Touschek wrote to his father on December 24, 1958, ``{\it Without him for me physics is only half interesting...}''.

 Touschek’s interaction with Pauli and other theorists (such as Charles Enz, Gerhart L\"uders, Markus Fierz, Kurt Symanzik, Luigi Radicati, G. Morpurgo, M. Cini) was instrumental in the development of his ideas on QED and discrete symmetries, as well as in stimulating his own reflections on the CPT theorem -- the solid conceptual base for AdA.\footnote{In this regard, see correspondence of the period in his papers at `Edoardo Amaldi Archives' in Rome, and published papers. G. Morpurgo, B. Touschek, L. Radicati. 1954. \href{https://doi.org/10.1007/BF02781835}{On time reversal}; G. Morpurgo, B. Touschek. 1955. \href{https://doi.org/10.1007/BF02731764}{Remarks on time reversal}; G. Morpurgo, B. Touschek. 1955. \href{https://doi.org/10.1007/BF02731422}{Space and time reflection of observable and non-observable quantities in field theory}; G. Morpurgo, B. Touschek. 1956. \href{https://doi.org/10.1007/BF02747964}{Space and time reflection in quantum field theory}; B. Touschek. 1957. \href{https://doi.org/10.1007/BF02835605}{Parity conservation and the mass of the neutrino}; B. Touschek. 1957. \href{https://doi.org/10.1007/BF02731633}{The mass of the neutrino and the non-conservation of parity}; L. Radicati, B. Touschek. 1957. \href{https://doi.org/10.1007/BF02856061}{On the equivalence theorem for the massless neutrino}; M. Cini, B. Touschek. 1958. \href{https://doi.org/10.1007/BF02747708}{The relativistic limit of the theory of spin 1/2 particles}; B. Touschek. 1958. \href{https://doi.org/10.1007/BF02828864}{The symmetry properties of Fermi Dirac fields}; B. Touschek. 1959. \href{https://doi.org/10.1007/BF02732949}{A note on the Pauli trasformation}. This sequence of papers was followed by the first paper on the AdA project: C. Bernardini, G.F. Corazza, G. Ghigo, B. Touschek. 1960. \href{https://doi.org/10.1007/BF02733192}{The Frascati storage ring}.} 
 As he himself recalled, referring to the late 1950s, the years ushering the transition to a new phase of his scientific adventure:
 
 \begin{quote}
 {\it At the time I felt rather exhausted
from an overdose of work which I had been trying to perform in the most abstract
field of theoretical research: the discussion of symmetries which had been opened up by
the discovery of the breakdown of one of them, parity, by Lee and Yang. I therefore
wanted to get my feet out of the clouds and onto the ground again, touch things (provided
there was no high tension on them) and take them apart and get back to what I
thought I really understood: elementary physics.}\footnote{B. Touschek, ``AdA and Adone are storage rings'', manuscript in Bruno Touschek papers, Box 11, Folder 3.92.4, p. 7.}
 \end{quote}
 
 Touschek's conceptual leap into the new world of matter-antimatter physics was well expressed by  Rubbia himself [Rubbia 2004, 59]:
 \begin{quote}
{\it \dots in his mind electron-positron collisions were nothing else than the way of realizing in practice the idea of symmetry between matter and antimatter, in the deep sense of the Dirac equation \dots}
\end{quote}
 
The analysis of Touschek's scientific life in the 1950s is thus to be pursued as one of the main keys to a deeper understanding of all the implications of his unique path towards what became a standard practice: using matter-antimatter annihilations to probe the ultimate nature of the basic building blocks of the Universe and their interactions.

\begin{figure}[h]
\includegraphics[width=8.7cm]{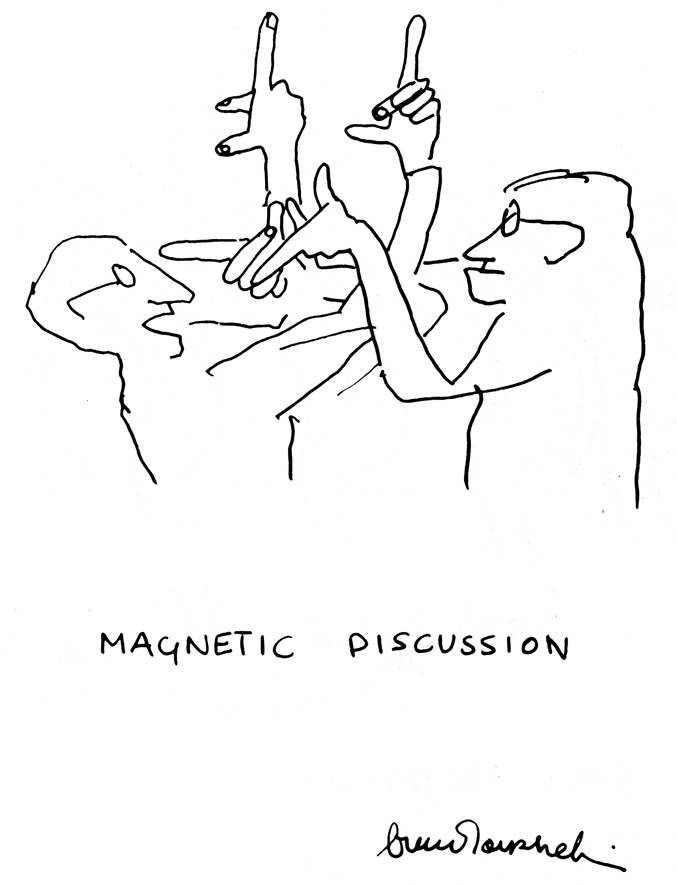}
\centering
\caption{Touschek's most celebrated drawing,  where Touschek is expressing with his usual pictorial effectiveness discussions about which were the electrons ad which were the positrons circulating in opposite directions in AdA. Touschek considered such a dilemma definitely irrelevant, as he saw the whole question just as an obvious `manifestation of CPT'. Courtesy of Francis Touschek.}

\end{figure}

\section*{Bibliography on Bruno Touschek and AdA}

The following fairly complete bibliography in chronological order is intended to be a useful tool to learn about the life and science of this extraordinary protagonist of 20th century physics. 

Whenever possible, a link to each reference has been provided, including the two docu-films, which are both accessible and visible online.\footnote{All links have been last accessed on 30/10/2021. In case of any problem to retrieve articles, please ask the author of the present contribution.}

\vskip 0.5cm

\indent 
C. Bernardini. 1978. Storia di AdA, {\it Scientia} 113: 27-38.
\vskip 0,3 cm
E. Amaldi.  1981. \href{http://cdsweb.cern.ch/record/135949/files/CERN-81-19.pdf}{\it The Bruno Touschek Legacy (Vienna 1921 -- Innsbruck 1978)}.  CERN Yellow Reports, No. 81-19.
\vskip 0,3 cm
E. Amaldi. 1982.  L'eredità di Bruno Touschek, {\it Quaderni del Giornale di Fisica} 5  (7).
\vskip 0,3 cm
C. Bernardini. 1986. Storia dell'anello AdA, {\it Il Nuovo Saggiatore} 27 (6): 23-33.
\vskip 0,3 cm
G. Battimelli, M. De Maria, G. Paoloni. 1989.  {\it Le carte di Bruno Touschek}, Università Sapienza, Roma.
\vskip 0,3 cm
F. Amman. 1989. The early times of electron colliders, in  M. De Maria, M. Grilli, F. Sebastiani (eds.),  {\it The Restructuring of Physical Sciences in Europe and the United States, 1945-1960}, World Scientific, Singapore, pp. 449-476.
\vskip 0,3 cm
C. Bernardini. 1989. AdA: the smallest $e^+e^-$ ring, in M. De Maria, M. Grilli, F. Sebastiani (eds.),  {\it The Restructuring of Physical Sciences in Europe and the United States, 1945-1960}, World Scientific, Singapore, pp. 444-448.
\vskip 0,3 cm
C. Bernardini. 1991. From the Frascati Electron Synchrotron to Adone, in Bacci et al. (eds.), {\it Present and Future of Collider Physics: Conference in honour of Giorgio Salvini's 70th birthday}, SIF, Bologna, pp. 3-15.
\vskip 0,3 cm
R. Wider\o e. 1994. \href{https://doi.org/10.1007/978-3-663-05244-9}{\it The Infancy of particle accelerators. Life and work of Rolf Wider\o e}, edited by P. Waloschek. Vieweg+Teubner Verlag, Braunschweig, Germany.
\vskip 0,3 cm
C. Bernardini. 1997. Bruno Touschek and AdA, in V. Valente, {\it Adone, a milestone on the particle way}, INFN, Frascati, pp. 1-21.
\vskip 0,3 cm
F. Amman. 1997. ADONE and the International Collaboration in the $e^+e^-$ Storage Rings Development, in V. Valente, {\it Adone, a milestone on the particle way}, INFN, Frascati, pp. 25-55.
\vskip 0,3 cm
N. Cabibbo. 1997. $e^+e^-$  Physics -- a View from Frascati in 1960's, in V. Valente, {\it Adone, a milestone on the particle way}, INFN, Frascati, pp. 219-225.
\vskip 0,3 cm
C. Bernardini. 1997. AdA e Frascati, in {\it Quark 2000}, Le Scienze, pp. 58-65.
\vskip 0,3 cm
G. Pancheri. May 1998. \href{https://cds.cern.ch/record/718672?ln=en}{Il grande Touschek, `Signore degli anelli'}. {\it  Corriere della Sera}.
\vskip 0,3 cm
G. Salvini. 1998. The Frascati decision and the AdA proposal, in G. Isidori (ed.), {\it Bruno Touschek and the birth of $e^+e^-$  physics}, Frascati Physics Series Vol. XIII, pp. 1-8.
\vskip 0,3 cm
C. Bernardini. 1998. Remembering Bruno Touschek, his work and personality, in G. Isidori (ed.), {\it Bruno Touschek and the birth of $e^+e^-$  physics}, Frascati Physics Series Vol. XIII, pp. 9-16.
\vskip 0,3 cm
J. Ha\"issinski. 1998. {From AdA to ACO, Reminiscences of Bruno Touschek}, in G. Isidori (ed.), {\it Bruno Touschek and the birth of $e^+e^-$ physics}, Frascati Physics Series Vol. XIII,  pp. 17-31.
\vskip 0,3 cm
M. Testa. 1998. The Adone results and the development of the Quark Parton Model, in G. Isidori (ed.), {\it Bruno Touschek and the birth of $e^+e^-$  physics}, Frascati Physics Series Vol. XIII, pp. 33-38.
\vskip 0,3 cm
E. Picasso. 1998. Electron-Positron storage rings from AdA to LEP, in G. Isidori (ed.), {\it Bruno Touschek and the birth of $e^+e^-$  physics}, Frascati Physics Series Vol. XIII, pp. 39-50.
\vskip 0,3 cm
CERN Courier, Feature. February 1999. \href{https://cds.cern.ch/record/1732884}{Electron-positron pioneer}. {\it CERN Courier} 39 (1): 17.
\vskip 0,3 cm
C. Bernardini. 1999. Bruno Touschek, {\it Il Nuovo Saggiatore} 15 (3-4): 29-33.
\vskip 0,3 cm
C. Bernardini. 2002. \href{http://www.sisfa.org/wp-content/uploads/2013/03/002-BERNARDINI-DEFINITIVO.pdf}{La nascita degli anelli di accumulazione per elettroni e positroni}, in M. Leone, A. Paoletti, N. Robotti (eds.). {\it Atti del XXII Congresso Nazionale di Storia della Fisica e dell'Astronomia}, pp. 27-36.
\vskip 0,3 cm
E. Agapito, L. Bonolis. 2004.  \href{https://www.youtube.com/watch?v=R2YOjnUGaNY}{Docu-Film {\it Bruno Touschek and the Art of Physics}}. Original idea and screenplay: Luisa Bonolis. Directed by Enrico Agapito.
\vskip 0,3 cm
C. Bernardini. 2004. \href{https://link.springer.com/article/10.1007\%2Fs00016-003-0202-y}{Ada: the first electron-positron collider}. {\it Physics in Perspective} 6: 156-183.
\vskip 0,3 cm
M. Greco, G. Pancheri (eds.). 2004. \href{http://www.lnf.infn.it/sis/frascatiseries/Volume33/volume33.pdf}{{\it Bruno Touschek Memorial Lectures}}.   Frascati Physics Series Vol. XXXIII.
\vskip 0,3 cm
C. Rubbia. 2004. \href{http://www.lnf.infn.it/sis/frascatiseries/Volume33/volume33.pdf}{The role of Bruno Touschek in proton-antiproton collider physics}, in M. Greco, G. Pancheri (eds.), {\it Bruno Touschek Memorial Lectures},   Frascati Physics Series Vol. XXXIII, pp. 57-60.
\vskip 0,3 cm
G. Salvini. 2004. \href{http://www.lnf.infn.it/sis/frascatiseries/Volume33/volume33.pdf}{Matter-antimatter collisions from Frascati to the outside world}, in M. Greco, G. Pancheri (eds.), {\it Bruno Touschek Memorial Lectures},   Frascati Physics Series Vol. XXXIII, pp. 61-68.
\vskip 0,3 cm
R. Gatto. 2004. \href{http://www.lnf.infn.it/sis/frascatiseries/Volume33/volume33.pdf}{Memories of Bruno Touschek}, in M. Greco, G. Pancheri (eds.), {\it Bruno Touschek Memorial Lectures},   Frascati Physics Series Vol. XXXIII, pp. 69-76.
\vskip 0,3 cm
C. Bernardini. 2004. \href{http://www.lnf.infn.it/sis/frascatiseries/Volume33/volume33.pdf}{The AdA Storage Ring}, in M. Greco, G. Pancheri (eds.), {\it Bruno Touschek Memorial Lectures},   Frascati Physics Series Vol. XXXIII, pp. 77-79.
\vskip 0,3 cm
G. Morpurgo. 2004. \href{http://www.lnf.infn.it/sis/frascatiseries/Volume33/volume33.pdf}{My work with Bruno Touschek}, in M. Greco, G. Pancheri (eds.), {\it Bruno Touschek Memorial Lectures},   Frascati Physics Series Vol. XXXIII, pp. 80-88.
\vskip 0,3 cm
U. Amaldi. 2004. \href{http://www.lnf.infn.it/sis/frascatiseries/Volume33/volume33.pdf}{Remembering Bruno Touschek}, in M. Greco, G. Pancheri (eds.), {\it Bruno Touschek Memorial Lectures},   Frascati Physics Series Vol. XXXIII, pp. 89-92.
\vskip 0,3 cm
G. Sacerdoti. 2004. \href{http://www.lnf.infn.it/sis/frascatiseries/Volume33/volume33.pdf}{Remembering Bruno Touschek}, in M. Greco, G. Pancheri (eds.), {\it Bruno Touschek Memorial Lectures},   Frascati Physics Series Vol. XXXIII, pp. 93-95.
\vskip 0,3 cm
F. Calogero. 2004. \href{http://www.lnf.infn.it/sis/frascatiseries/Volume33/volume33.pdf,}{Remembering Bruno Touschek}, in M. Greco, G. Pancheri (eds.),  {\it Bruno Touschek Memorial Lectures},   Frascati Physics Series Vol. XXXIII, p. 96.
\vskip 0,3 cm
P. Waloschek. 2004. \href{http://www.lnf.infn.it/sis/frascatiseries/Volume33/volume33.pdf}{Remembering Bruno Touschek}, in M. Greco, G. Pancheri (eds.),  {\it Bruno Touschek Memorial Lectures},   Frascati Physics Series Vol. XXXIII, pp. 97-98.
\vskip 0,3 cm
G. Rossi. 2004. \href{http://www.lnf.infn.it/sis/frascatiseries/Volume33/volume33.pdf}{Remembering Bruno Touschek}, in M. Greco, G. Pancheri (eds.), {\it Bruno Touschek Memorial Lectures},   Frascati Physics Series Vol. XXXIII, pp. 99-100.
\vskip 0,3 cm
G. Pancheri. 2004. \href{http://www.lnf.infn.it/sis/frascatiseries/Volume33/volume33.pdf}{Bruno Touschek and the Frascati Theory Group}, in M. Greco, G. Pancheri (eds.), {\it Bruno Touschek Memorial Lectures},   Frascati Physics Series  Vol. XXXIII,  pp. 101-104.
\vskip 0,3 cm
L. Bonolis. 2005a. \href{https://www.sif.it/riviste/sif/ncr/econtents/2005/028/11/article/0}{Bruno Touschek vs. machine builders: AdA, the first matter-antimatter collider}. {\it Rivista del Nuovo Cimento} 28 (11): 1-60. 
\vskip 0,3 cm
C. Bernardini. 2005. \href{http://www.analysis-online.net/wp-content/uploads/2013/03/bernardini_tous.pdf}{Bruno Touschek visto da vicino}, {\it Analysis} 4: 4-7. 
\vskip 0,3 cm
G. Pancheri. 2005. \href{http://www.analysis-online.net/wp-content/uploads/2013/03/pancheri_tous.pdf}{Bruno Touschek e la nascita della fisica $e^+e^-$: una storia europea}, {\it Analysis} 4: 8-15. 
\vskip 0,3 cm
L. Bonolis. 2005b. \href{http://www.analysis-online.net/wp-content/uploads/2013/03/bonolis_rivoluzione.pdf}{Una rivoluzione culturale nel mondo degli acceleratori di particelle: Bruno Touschek e il primo anello di collisione materia-antimateria},  {\it Analysis} 4: 16-32. 
\vskip 0,3 cm
V. Valente. 2007. La rivoluzione di AdA, in V. Valente, {\it Strada del Sincrotrone km 12}, Ch. 4.
\vskip 0,3 cm
L. Bonolis. 2007. Bruno Touschek e la genesi di un'idea: AdA, il primo anello di accumulazione per elettroni e positroni, in M. Leone et al. (eds.), {\it L'eredit\`a di Fermi e Majorana, e  altri temi}, Bibliopolis, pp. 217-222.
\vskip 0,3 cm
C. Bernardini. 2007. Bruno Touschek, in N. Koertge (ed. in chief), {\it New Dictionary of Scientific Biography}, New York, Charles Scribner's Sons, pp. 68-73.
\vskip 0,3 cm
M. Greco, G. Pancheri. 2008. \href{http://www.analysis-online.net/wp-content/uploads/2013/03/greco_pancheri.pdf}{Frascati e la fisica teorica: da AdA a Dafne}, {\it Analysis} 2+3: 21--38.
\vskip 0,3 cm
L. Bonolis. December 2008. Bruno Touschek, un mitteleuropeo ai Laboratori Nazionali di Frascati, {\it Sapere}:  6-13.
\vskip 0,3 cm
C. Bernardini. 2008. Bruno Touschek: pensare fisica in grande, {\it La Fisica nella Scuola} 41 (4): 153-159.
\vskip 0,3 cm
L. Bonolis, G. Pancheri. 2009. \href{https://higherlogicdownload.s3.amazonaws.com/APS/5850cbf5-d2ca-4fe3-9812-9b34f575294f/UploadedImages/Documents/dec09.pdf}{Bruno Touschek and the Birth of Electron-Positron Collisions}, {\it International Newsletter}, Forum on International Physics - The American Physical Society, December 31: 16-18. 
\vskip 0,3 cm
L. Bonolis, G. Pancheri. 2011. \href{http://dx.doi.org/10.1140/epjh/e2011-10044-1}{Bruno Touschek: particle physicist and father of the $e^+e^-$ collider}, {\it European Physical Journal H} 36 (1): 1-61. 
\vskip 0,3 cm
P. Waloschek. 2012. \href{http://www-library.desy.de/preparch/books/death-rays.pdf}{\it Death-Rays as Life-Savers in the Third Reich}. DESY.
\vskip 0,3 cm
E. Agapito, L. Bonolis, G. Pancheri. 2013. \href{http://www.lnf.infn.it/edu/materiale/video/AdA_in_Orsay.mp4}{Docu-Film {\it Touschek with AdA in Orsay}}. 
Original idea and screenplay: Luisa Bonolis and Giulia Pancheri.
Directed by Enrico Agapito.
\vskip 0,3 cm
CERN Courier, Feature. 22nd January 2014. \href{https://cerncourier.com/a/ada-the-small-machine-that-made-a-big-impact/}{AdA – the small machine that made a big impact. The first electron-positron collisions at a storage ring were observed 50 years ago.}
\vskip 0,3 cm
E. Iarocci. 2015. \href{https://www.ilnuovosaggiatore.sif.it/download/14}{AdA: il successo di un'idea}, {\it Il Nuovo Saggiatore} 27 (1-2): 17-28. 
\vskip 0,3 cm
C. Bernardini, G. Pancheri, C. Pellegrini. 2015. \href{https://doi.org/10.1142/S1793626815300133}{Bruno Touschek: From Betatrons to Electron--Positron Colliders}, {\it Reviews of Accelerator Science and Technology} 8: 269-290. \href{https://arXiv.org/abs/1510.00933}{Arxiv:1510.00933}.
\vskip 0,3 cm
G. Pancheri, L. Bonolis. 2017. \href{https://arxiv.org/abs/1710.09003}{The path to high-energy electron-positron colliders: from Widerøe's betatron to Touschek's AdA and to LEP}. arXiv:1710.09003.
\vskip 0,3 cm
L. Bonolis, G. Pancheri. 2018. \href{https://arxiv.org/abs/1805.09434}{Bruno Touschek and AdA: from Frascati to Orsay. In memory of Bruno Touschek, who passed away 40 years ago, on May 25th, 1978}. arXiv:1805.09434. Report INFN -- 18-05/LNF, Istituto Nazionale di Fisica Nucleare, Laboratori Nazionali di Frascati, May 25, 2018. 
\vskip 0,3 cm
G. Pancheri, L. Bonolis. 2018. \href{https://arxiv.org/abs/1812.11847}{Touschek with AdA in Orsay and the first direct observation of electron-positron collisions}.  arXiv:1812.11847. Report INFN -- 18-12/LNF, Istituto Nazionale di Fisica Nucleare, Laboratori Nazionali di Frascati, December 31, 2018.
\vskip 0,3 cm
L. Bonolis, G. Pancheri. 2019a. \href{https://www.bshs.org.uk/wp-content/uploads/Viewpoint_118_Web.pdf}{A Tale of Two Scientists and the Development of Particle Colliders}. {\it Viewpoint} 118: 7-9. 
\vskip 0,3 cm
L. Bonolis, G. Pancheri. 2019b. \href{https://arxiv.org/abs/1910.09075}{Bruno Touschek in Germany after the War: 1945-46}. arXiv:1910.09075. Report INFN -- 19-17/LNF, Istituto Nazionale di Fisica Nucleare, Laboratori Nazionali di Frascati, and MIT-CTP/5150, October 10, 2019. 
\vskip 0,3 cm
L. Bonolis, G. Pancheri. 2019c. \href{https://www.treccani.it/enciclopedia/bruno-touschek_\%28Dizionario-Biografico\%29/}{Touschek, Bruno}, in {\it Dizionario Biografico degli Italiani}, Vol. 96.
\vskip 0,3 cm
G. Pancheri, L. Bonolis. 2020. \href{https://arxiv.org/abs/2005.04942}{Bruno Touschek in Glasgow. The making of a theoretical physicist}. arXiv:2005.04942.
\vskip 1 cm

 \begin{figure}[h]
\includegraphics[width=15cm]{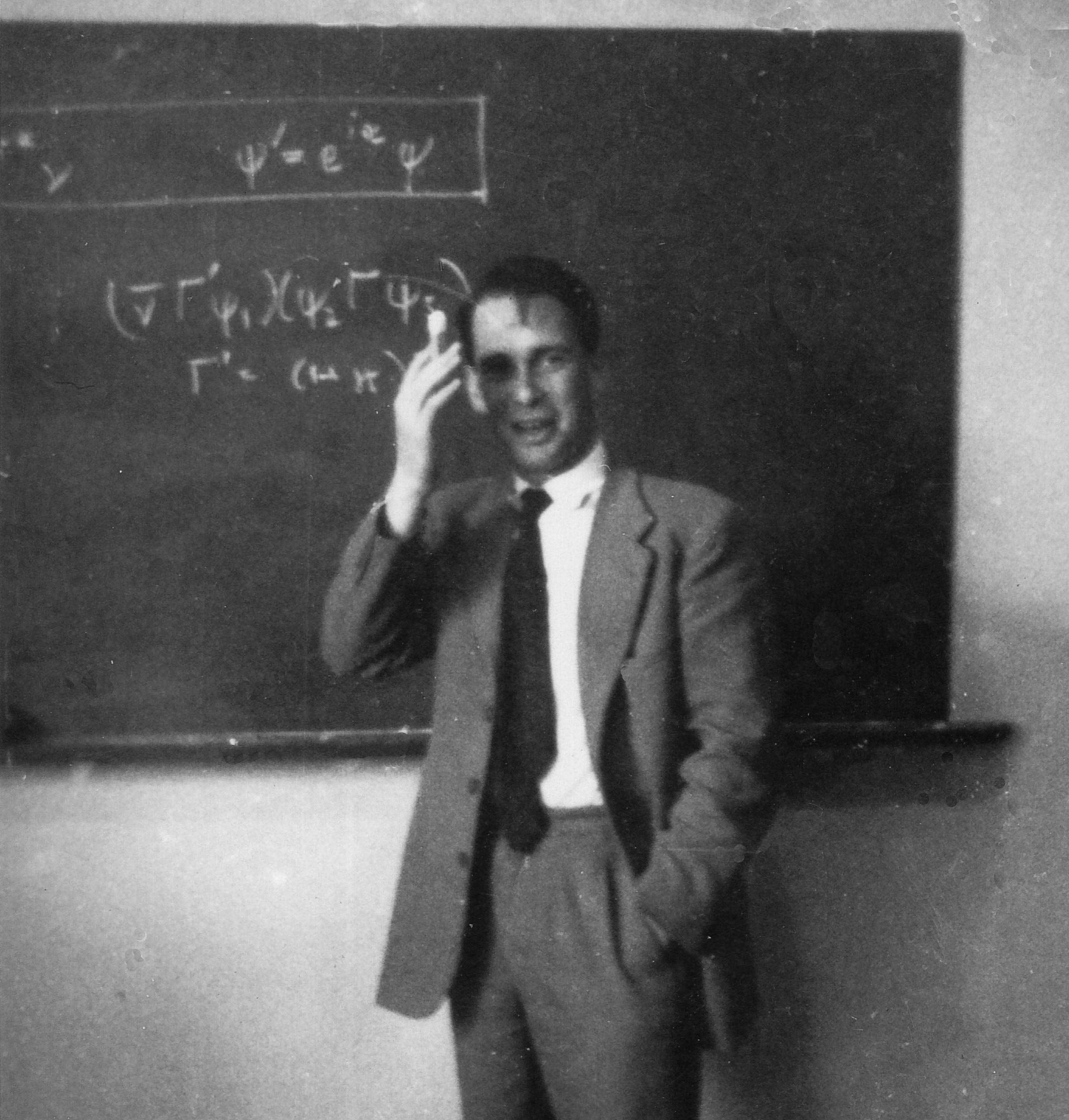}
\centering
\caption{Bruno Touschek in the 1960s. Courtesy of Francis Touschek.}
\end{figure}

\end{document}